\documentstyle[12pt,epsf,a4]{article}

\newcommand{\AmS}{{\protect\the\textfont2  
  A\kern-.1667em\lower.5ex\hbox{M}\kern-.125emS}}
\newcommand{\bq}{\begin{equation}}
\newcommand{\eq}{\end{equation}}
\newcommand{\beq}  {\begin{eqnarray}}
\newcommand{\eeq}  {\end{eqnarray}}
\newcommand{\rG}   {{\rm GUT}}
\newcommand{\MG}   {{\ifmmode M_\rG         \else $M_\rG$          \fi}}
\newcommand{\mb}   {{\ifmmode m_{b}         \else $m_{b}$          \fi}}
\newcommand{\mt}   {{\ifmmode m_{t}         \else $m_{t}$          \fi}}
\newcommand{\agut} {{\ifmmode \alpha_\rG    \else $\alpha_\rG$     \fi}}
\newcommand{\mgut} {{\ifmmode M_\rG         \else $M_\rG$          \fi}}
\newcommand{\mze}  {{\ifmmode m_0           \else $m_0$            \fi}}
\newcommand{\mha}  {{\ifmmode m_{1/2}       \else $m_{1/2}$        \fi}}
\newcommand{\tb}   {{\ifmmode \tan\beta     \else $\tan\beta$      \fi}}
\newcommand{\mz}   {{\ifmmode M_{Z}         \else $M_{Z}$          \fi}}
\newcommand{\ai}   {{\ifmmode \alpha_i      \else $\alpha_i$       \fi}}
\newcommand{\aii}  {{\ifmmode \alpha_i^{-1} \else $\alpha_i^{-1}$  \fi}}
\newcommand{\MSb}  {{\ifmmode \overline{\rm MS} \else
                             $\overline{\rm MS}$                   \fi}}
\newcommand{\DRb}  {{\ifmmode \overline{\rm DR} \else
      $\overline{\rm DR}$                   \fi}}
\newcommand{\DRbar}{{\ifmmode \overline{DR} \else $ \overline{DR}$ \fi}}
\newcommand{\msusy}{{\ifmmode M_{SUSY}      \else $M_{SUSY}$       \fi}}
\newcommand{\as}   {{\ifmmode \alpha_s      \else $\alpha_s$       \fi}}
\newcommand{\asmz} {{\ifmmode \alpha_s(M_Z) \else $\alpha_s(M_Z)$  \fi}}
\newcommand{\tal}  {{\ifmmode \tilde{\alpha} \else $\tilde{\alpha}$ \fi}}

\newcommand{\sws}  {{\ifmmode \;\sin^2\theta_W
                     \else    $\;\sin^{2}\theta_{W}$               \fi}}
\newcommand{\cws}  {{\ifmmode \;\cos^2\theta_W  
                     \else    $\;\cos^{2}\theta_{W}$               \fi}}
\newcommand{\sw}   {{\ifmmode\;\sin\theta_W\else $\sin\theta_{W}$  \fi}}
\newcommand{\cw}   {{\ifmmode\;\cos\theta_W\else $\;\cos\theta_{W}$\fi}}
\newcommand{\tw}   {{\ifmmode\;\tan\theta_W\else $\;\tan\theta_{W}$\fi}}
\newcommand{\bsg}  {{\ifmmode b\rightarrow s\gamma
                     \else $b\rightarrow s\gamma$ \fi}}
\newcommand{\Bbsg}  {{\ifmmode BR(\b\rightarrow s\gamma)
\else $BR(b\rightarrow s\gamma)$ \fi}}

\newcommand{\rPL}  {{\rm Planck}}
\newcommand{\mplanck} {{\ifmmode M_\rPL         \else $M_\rPL$          \fi}}
\newcommand{\rST}  {{\rm SO(10)}}
\newcommand{\msoten} {{\ifmmode M_\rST         \else $M_\rST$          \fi}}

\def\be{\begin{equation}}
\def\ee{\end{equation}}
\def\bea{\begin{eqnarray}}
\def\eea{\end{eqnarray}}

\begin{document}
\begin{center}
  {\large\bf Updated Combined Fit of Low Energy Constraints \\[3mm]
            to Minimal Supersymmetry} \\[0.8cm]

  {\bf W.~de Boer, R.~Ehret, J.~Lautenbacher} \\[5mm]
  {\it Institut f\"ur Experimentelle Kernphysik, University of Karlsruhe \\
       Postfach 6980, D-76128 Karlsruhe, Germany} \\[0.8cm]

  {\bf A.V.~Gladyshev, D.I.~Kazakov} \\[5mm]

{\it Bogoliubov Laboratory of Theoretical Physics,
Joint Institute for Nuclear Research, \\
141 980 Dubna, Moscow Region, Russian Federation}
\end{center}

\begin{abstract}
The new precise LEP measurements of $\alpha_S$ and $\sin^2\theta_W$
as well as the new LEP II mass limits for supersymmetric particles
and new calculations for the radiative (penguin) decay of the $b$-quark 
into $s\gamma$ allow a further restriction in the parameter 
space of the Constrained Minimal Supersymmetric
Standard Model (CMSSM). 
\end{abstract}

\section{Introduction}
  \begin{figure}[ht]
\vspace{-0.3cm}
    \begin{center}
    \leavevmode
    \epsfxsize=9cm
    \epsffile{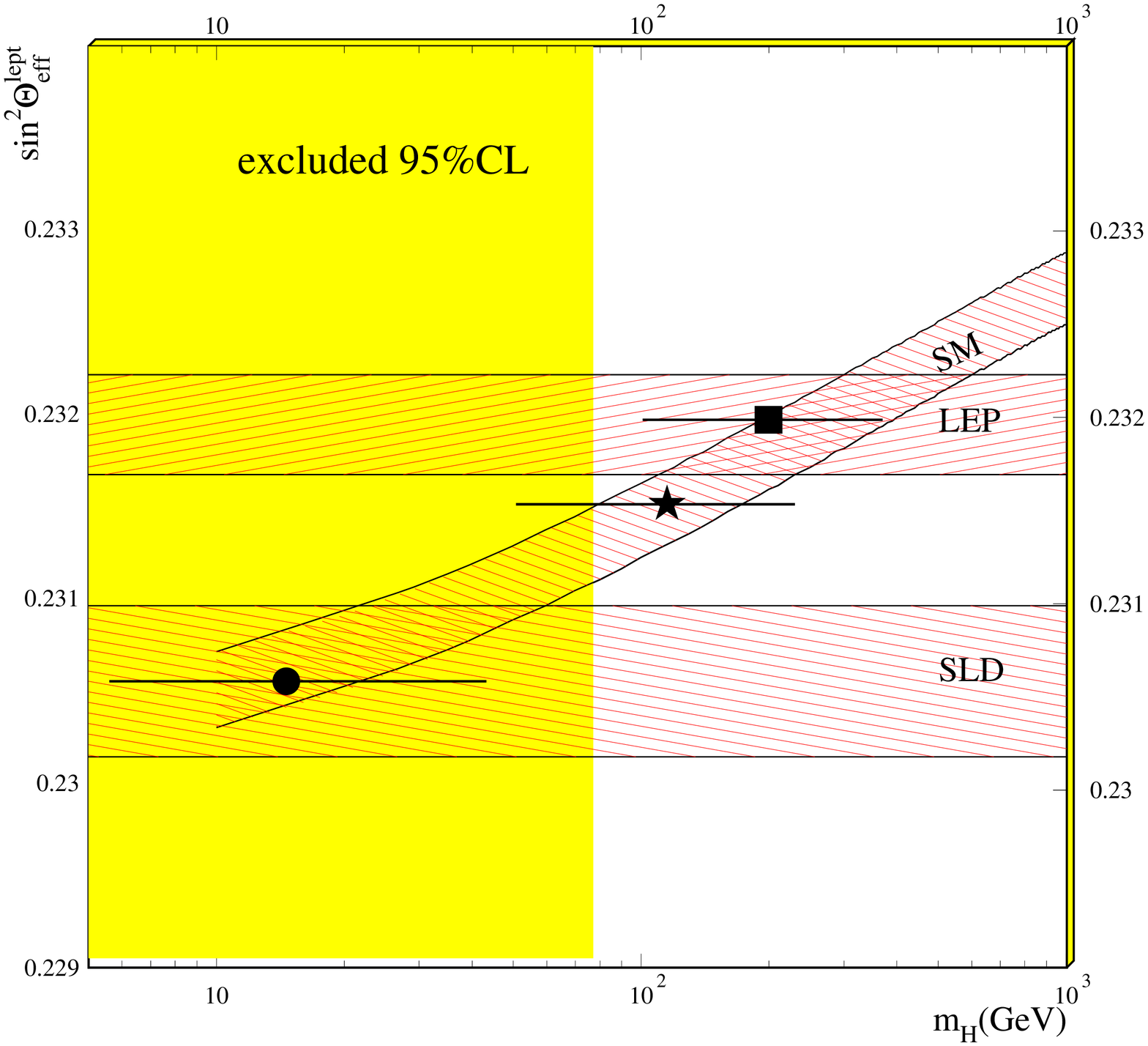}
\end{center}
\vspace{-0.5cm}
  \caption{\label{f2}
         The electroweak mixing angle versus Higgs mass.
         The diagonal band corresponds to the SM prediction; the 
         width of the band is determined by the uncertainty in the top mass.
         The point at $m_h=14.6^{+28.6}_{-14.6}$ GeV corresponds  to
         the SLC value of the mixing angle, while the point
         labeled LEP corresponds to the LEP value and the intermediate
         point corresponds to the combined data.
      }
\vspace{-0.4cm}
 \end{figure}
  \begin{figure}[ht]
\vspace{-1.7cm}
    \begin{center}
    \leavevmode
    \epsfxsize=9cm
    \epsffile{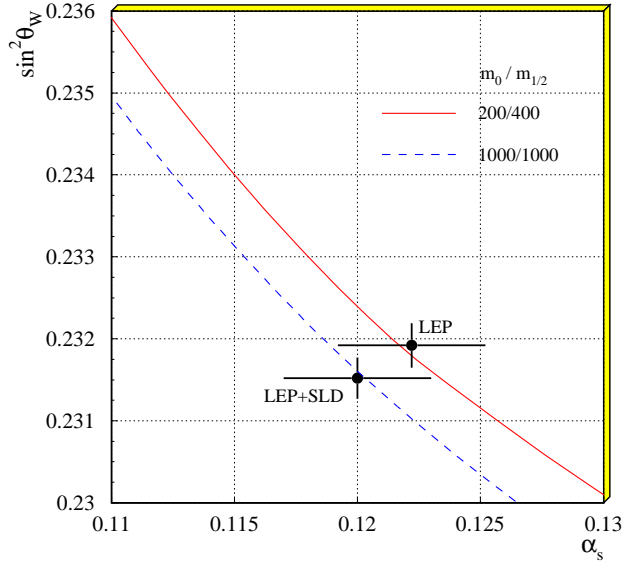}
\end{center}
\vspace{-2.5cm}
  \caption{\label{f3} The values of $\as$ and $\sin^2\Theta$
    yiel\-ding unification for $m_0=200,$ $ m_{1/2}$=400 
       and $m_0=1000, m_{1/2}=1000$ GeV.
       The upper curve fits better the LEP couplings, while the
       lower curve fits better the combined data of LEP and SLC.
      }
\vspace{-0.3cm}
  \end{figure}
  \begin{figure}[ht]
\vspace{-1cm}
    \begin{center}
    \leavevmode
    \epsfxsize=8cm
    \epsffile{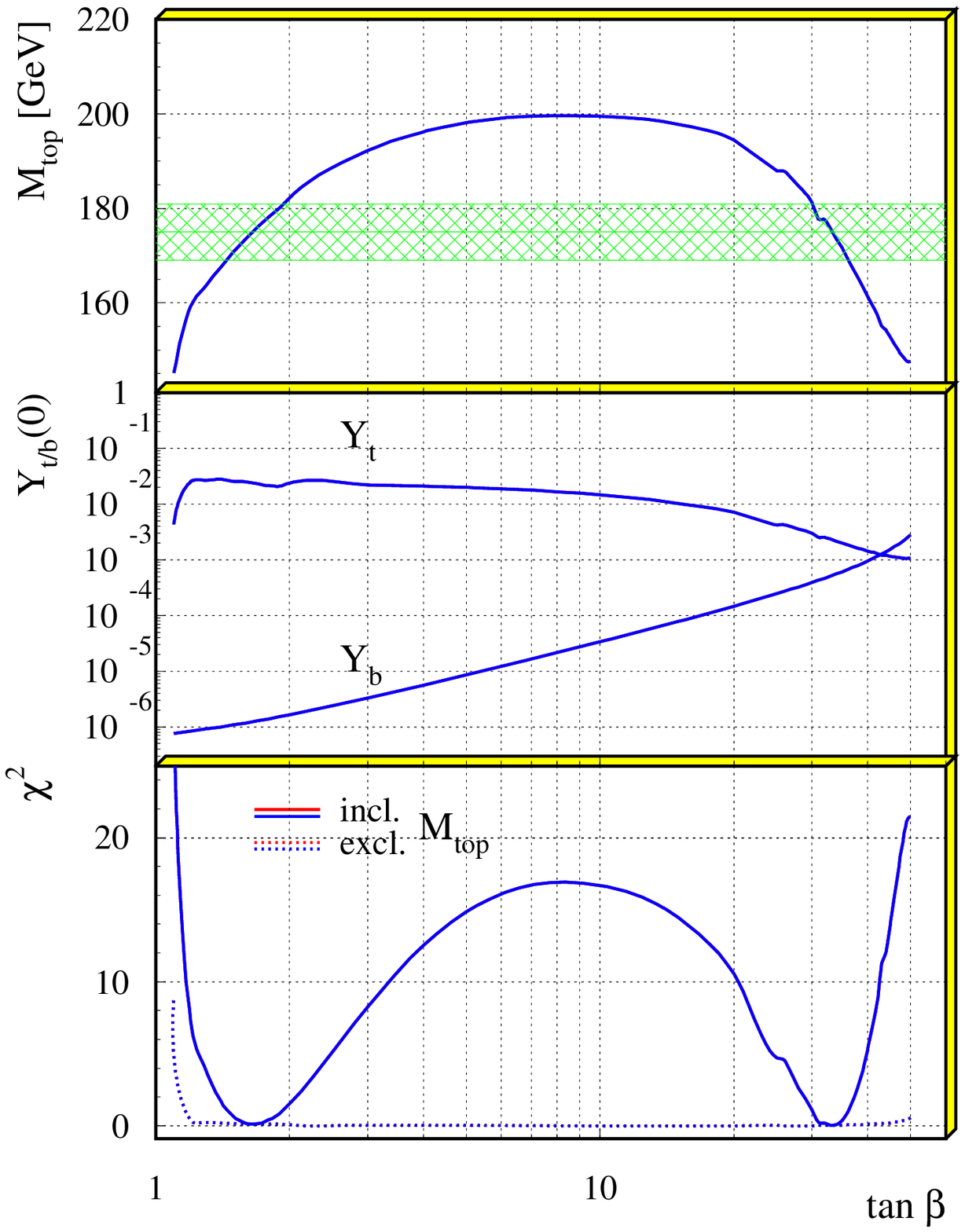}
\end{center}
\vspace{-1.5cm}
  \caption{\label{f4}The top quark mass as function of $\tb$ (top) 
      for values of $\mze,\mha~\approx 1 $ TeV.  The curve
      is hardly changed for lower SUSY masses.
      The middle part shows the corresponding values of the Yukawa
      coupling at the GUT scale and the lower part the
      $\chi^2$ values.
       If the top constraint ($\mt=175\pm6$, horizontal band) 
      is not applied, all values of $\tb$ between 1.2 and 50 
       are allowed
      (thin dotted lines at the bottom), but if the top mass 
      is constrained to the experimental value, only the regions
      $1\le\tb\le3$ and $20\le \tb\le 40$ are allowed.
      }
 \end{figure}
\noindent In this paper we repeat our previous fit of the supergravity parameters\cite{wezp}
with the new input data from LEP concerning the coupling constants
and new higher order calculations for the   
$b\rightarrow s\gamma$  decay rate branching ratio~\cite{misiak}.
 The latter indicate that next to leading log (QCD) corrections
increase the SM value by about 10\%. This can be simulated 
in the lowest level calculation by choosing a renormalization 
scale $\mu=0.57m_b$,
which is  done in the following.
In addition, a  new preliminary measurement of the 
$b\rightarrow s\gamma$ rate
has been given by the ALEPH Coll.\cite{aleph}:
 $BR=(3.38\pm0.74\pm0.85)*10^{-4},$
which is slightly higher than the  previously measured value by the 
CLEO Coll.\cite{cleo}:
$BR=(2.32\pm0.57\pm0.35)*10^{-4}.$
Both are consistent with the SM expectation 
$BR=(3.52\pm0.33)*10^{-4}.$
This value was calculated for $\as=0.122$, which is the 
best fit value to the electroweak data from LEP, thus excluding 
the SLC value of
$\sin^2\Theta_{eff}^{lept}$, which is inconsistent
with the present Higgs limit of 79 GeV\cite{janot}, as shown in fig. \ref{f2}.
It remains to be seen, if the SLC value is a  statistical
fluctuation or due to a systematic uncertainty. For the moment we have taken
the LEP values for the coupling constants 
($\as=0.122\pm0.003$ and $\sin^2\Theta_{\overline{MS}}=0.2319\pm0.00029$), 
which are slightly higher than the fit values from the combined LEP and SLC
data ($\as=0.120\pm0.003$ and $\sin^2\Theta_{\overline{MS}}=0.2315\pm0.0004$)
and lead to different unification conditions, as shown in fig. \ref{f3}.
In addition to the statistical errors on $\sin^2\Theta_{\overline{MS}}$
we added in quadrature the error from the uncertainty in the
QED coupling constant at $M_Z$, which is 0.0026 and aris\-es
mainly from the uncertainty of the hadronic vacuum polarization.
We did not use the world average of $\as=0.118$, but used only
the LEP value, for which the 3rd order QCD corrections have been calculated
and found to be small, so the uncertainties from the higher order corrections
are small.

In addition LEP has provided a new chargino mass limit around 90 GeV 
in case of a gaugino-like chargino and a heavy sneutrino~\cite{LEP}, 
which is exactly the MSSM case considered he\-re:
($\mu > m_{1/2}$ and $m_{\tilde{l}}> 100$ GeV).  
The  Higgs mass limit is now 79 GeV at 95\% C.L. 
for the SM Higgs~\cite{janot} , which  corresponds to the CMSSM case too, 
since  all the other Higgses are heavy and decouple in the CMSSM limit
considered here.
\section{Results}
The fitted supergravity parameters  are 
mainly sensitive to the following input data:
The GUT scale \mgut~ and coupling constant \agut~
 are  determined from gauge coupling unification,
 the Yukawa couplings $Y_t$, $Y_b$, $Y_\tau$ at the GUT scale 
 from the masses of the  third generation, 
$\mu$ from electroweak symmetry breaking  and
$\tb$ from $b\tau$-unification. Of course, in a $\chi^2$-fit all parameters
are determined simultaneously, thus taking all correlations into account.
Since $m_0$ and $m_{1/2}$ 
enter in all observables, and are strongly correlated, we 
perform the fit for all combinations of  $m_0$ and $m_{1/2}$
 between 100 GeV and 1 TeV in steps of 100 GeV.
%
%
%
  \begin{figure}[t]
    \vspace{-0.51cm}
\begin{center}
    \leavevmode
    \epsfxsize=9.0cm
    \epsffile{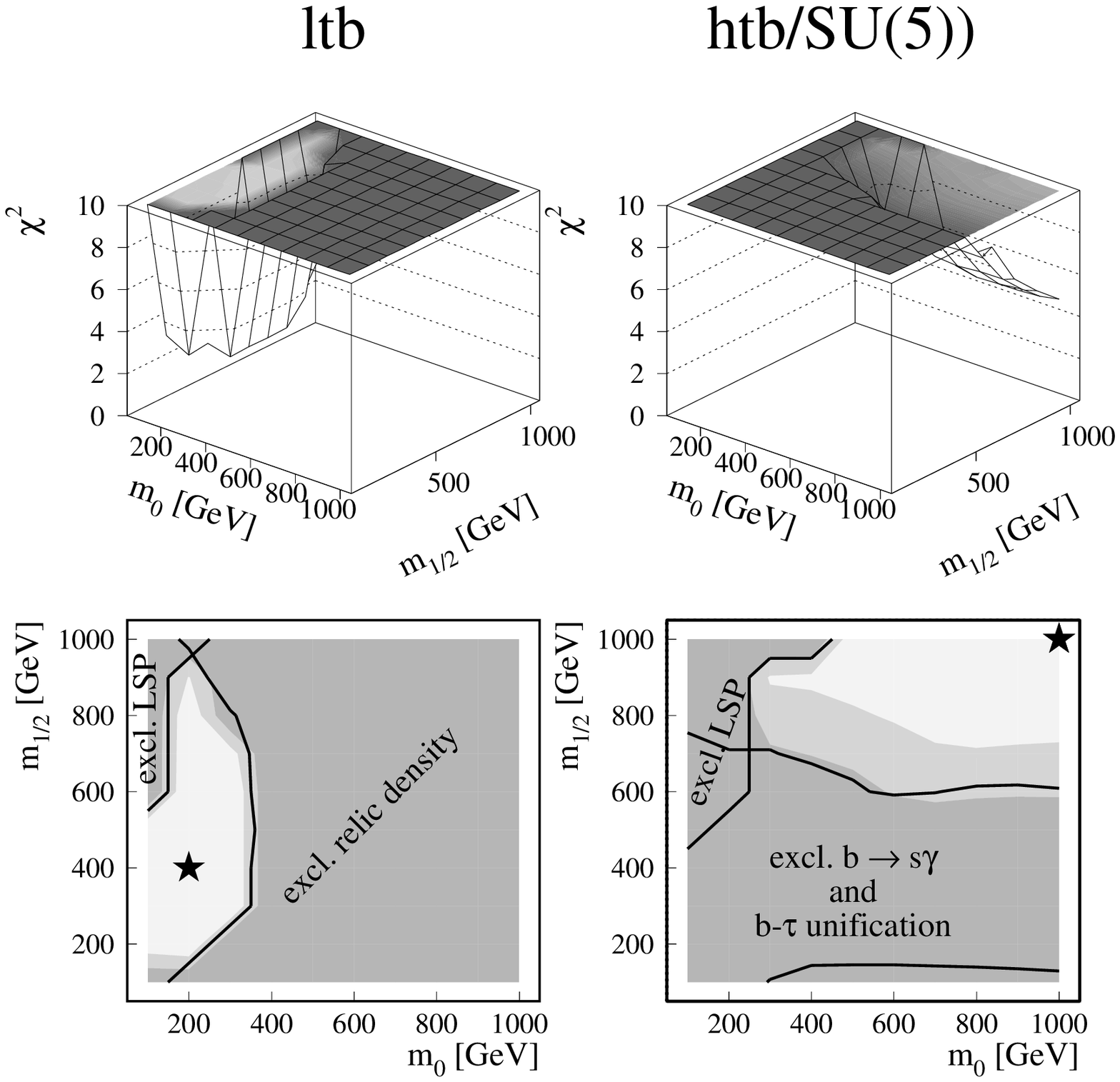}
\end{center}
\vspace{-0.8cm}
\caption[]{\label{f5}The  $\chi^2$-distribution for 
  the low and high $\tb$ solutions.  The different shades
   in the projections  indicate
  steps of $\Delta\chi^2 = 1$, so basically only the light shaded
  region is allowed.
      The stars indicate the optimum solution. 
      Contours enclose domains excluded by the particular
      constraints used in the analysis.
}
\end{figure}

%
%
  \begin{figure}[htb]
    \vspace{-0.51cm}
\begin{center}
    \leavevmode
    \epsfxsize=9.0cm
    \epsffile{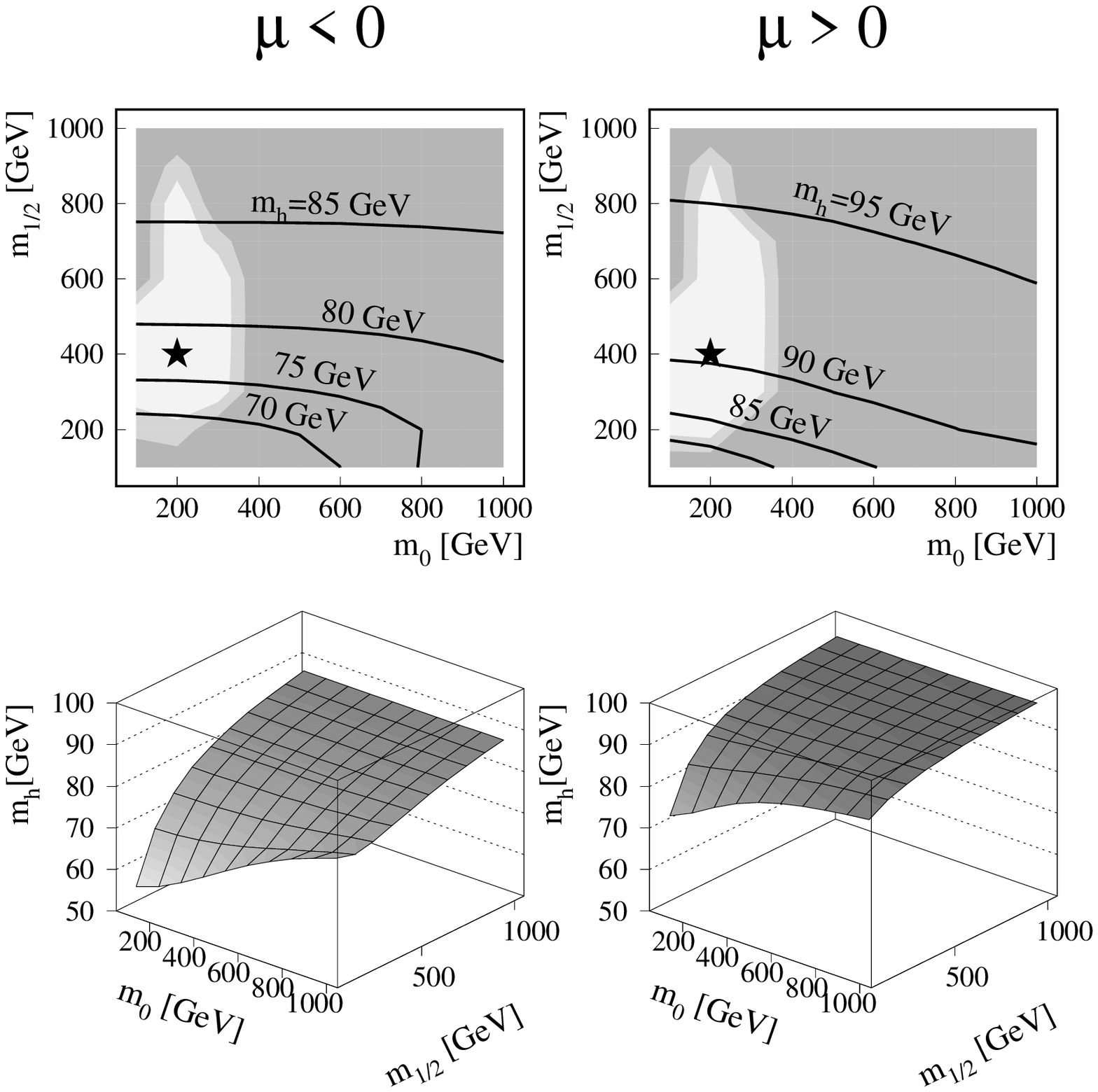}
\end{center}
\vspace{-0.8cm}
\caption[]{\label{f6}Contours of the  Higgs mass (solid lines) 
       in the $m_0,m_{1/2}$ plane (above)  and the Higgs masses (below) for both sign of $\mu$  for the low $\tb$ solution $\tb=1.65$
and $m_t=175$ GeV.
}
\end{figure}

The most  restrictive constraints are
the gauge
coupling constant unification and the requirement that the
unification scale has to be above $10^{15}$ GeV from the proton
lifetime limits, assuming decay via s-channel exchange of heavy
gauge bosons. They exclude the SM as well
as many other models.

The requirement of bottom-tau Yu\-ka\-wa coupling unification strongly
restricts the possible solutions in the $\mt$ versus $\tb$ plane.
For $m_t=175.6$ $\pm5.5$ GeV only two regions of $\tb$ give an
acceptable $\chi^2$ fit, as shown in 
fig.~\ref{f4}. $Y_t$ is left free independent of $Y_b=Y_\tau$.

In fig.~\ref{f5} the total $\chi^2$ distribution is shown as a
function of $\mze$ and $\mha$ for the two values of $\tb$ determined
above. One observes  minima at $\mze$, $\mha$ around (200,400) and
(1000,1000), as indicated by the stars. 
The contours in the lower part show the regions excluded by the
different constraints used in the analysis.

The requirement that the LSP is
  neutral excludes the regions with small $m_0$ and relatively large
  $m_{1/2}$, since in this case one of the scalar staus becomes the LSP
  after mixing via the off-diagonal elements in the mass matrix. 
  The LSP constraint is especially effective at
  the high \tb region, since the off-diagonal element in the stau mass matrix 
  is proportional
  to $A_t m_0 - \mu\tan\beta$.

  The \bsg rate is too large in
  most of the parameter region for large \tb, 
at least if one requires $b-\tau$ unification. Both $m_b$ and $b\to s\gamma$
have loop corrections proportional to $\mu\tan\beta$ and it is difficult to get
both constraints fulfilled simultaneously unless $\mu$ can be choosen to be small, which
can only be obtained for non-unified Higgs masses at the GUT scale.
Alternatively, one has to give up $b\tau$ unification, in which 
case a much larger region for high $\tb$ is allowed.

  The long lifetime of the universe requires a  mass density below the
  cri\-tical density, else the overclosed universe would have collapsed long ago.
  This requires that the selectron is sufficiently light for a fast
  annihilation through t-channel selectron exchange.
   For large $\tb$ the  
  Higgsino admixture becomes larger,  which leads to an enhancement of
  $\tilde\chi^0-\tilde\chi^0$ annihilation via the s-channel Z boson
  exchange, thus reducing the relic density.  As a result, in the
  large $\tb$ case the constraint $\Omega h_0^2 < 1$ is almost always
  satisfied unlike in the case of low $\tb$.

\section{Summary}
It is shown that the CMSSM can fit simultaneously the constraints from
\begin{itemize}
\item gauge couplings unification;
\item $b-\tau$ Yukawa couplings unification;
\item $b\rightarrow s\gamma$ rate;
\item radiative EWSB;
\item life time of the universe.
\end{itemize}
For the high $\tan\beta$ scenario the present limits of 
LEP~\cite{LEP} on the chargi\-no mass and the more 
precise NLO calculations of \bsg decay rate together with the measurements from 
ALEPH and CLEO 
leave only the small upper right corner region in parameter space
($m_0>500, m_{1/2}>500$ GeV) (see fig. \ref{f5}), where all squark  masses are above 1 TeV.

For the low $\tb$ scenario the Higgs mass is below 100 (90) GeV for
$\mu >0 $ ($\mu <0$) and $m_t\le 181$ GeV.
The excluded regions in the ($m_0,m_{1/2})$ plane for a 
given Higgs mass are shown in fig. \ref{f6} for two different signs of $\mu$. 
Clearly the $\mu<0$ solution is largely excluded, since   the present LEP run at 183 GeV
did not observe a positive signal and 
preliminary limits above 85 GeV were quoted\cite{lepc}. For $\mu>0$
$m_{1/2}>135$ GeV is required from the gaugino-like chargino limit $m_{\chi}\ge 90$ GeV and 
$m_0<350$ GeV from the relic density constraint (see fig. \ref{f5}). 


\begin{thebibliography}{10}
%
 \bibitem{wezp} W. de Boer et al., {\em Z. Phys. {\bf C71} (1996) 415.}
\bibitem{misiak} K.Chetyrkin et al, hep-ph/9612313\\
 A. Buras et al., hep-ph/9707482;\\
C. Greub and T.Hurth, hep-ph/9708214.
\bibitem{aleph} ALEPH Coll., contr. to HEP 97.
\bibitem{cleo} R. Ammar et al.,CLEO-coll., {\em Phys. Rev. Lett. {\bf 74} (1995) 2885.}
\bibitem{janot} P. Janot,  these proceedings. 
\bibitem{LEP} Contributed papers from the LEP Coll. to HEP 97.
%
\bibitem{lepc} LEP results presented  at the public session of the 
LEPC meeting, CERN, Geneva, Nov. 11, 1997.
\end{thebibliography}
\end{document}